\documentclass{article}

\usepackage{epsf}
\usepackage{graphicx}

\begin{document}

\twocolumn

\begin{center}
{\Large \bf Circumstellar Magnetic Field Diagnostics from 
Line Polarization}\\
\vspace{1em}

{\it\bf R.\ Ignace$^1$ \& K.~G.\ Gayley$^2$}\\
\vspace{1em}

{\it $^1$Department of Physics, Astronomy, \& Geology, East Tennessee State
University, USA}\\
{\it $^2$Department of Physics \& Astronomy, University of Iowa, USA}

\end{center}


\section{Introduction}

Given that dynamically significant magnetic fields in at least some
massive stars have now been measured, our contribution addresses
the question, to what extent can fields be directly detected in
circumstellar gas?  The question speaks directly to the very interesting
topic of line-driving physics coupled with magnetized plasmas, and how
this coupling produces structure in the wind flow.  The major goal of
this effort is the hope of relating direct measurements of photospheric
magnetic fields in massive stars, for example via the methods of Donati
\& Cameron (1997), with direct measurements of the circumstellar
magnetic field from wind lines.  Aside from non-thermal emissions,
direct detection of magnetic fields derives from the Zeeman effect.
Already, Donati et~al.\ (2005) has reported the detection of circularly
polarized lines in the disk of FU~Ori, signifying that the time is ripe
for modeling diagnostics of circumstellar magnetic fields to help guide
observers in similar future searches.

We focus our attention on weak-field diagnostics.  These come in two main
types:  the Hanle effect, which pertains to coherence effects for linear
polarization from line scattering, and the weak longitudinal Zeeman
effect, which pertains to circular polarization in lines.

\section{The Hanle Effect for Winds}

The Hanle effect refers to how a magnetic field can alter the linear
polarization of a scattering line.  When the splitting of magnetic
sublevels by the Zeeman effect $\Delta \nu_Z$ remains comparable to
the natural width of those sublevels $\Delta \nu_N$, a situation of
quantum coherence exists.   Normally, in the absence of a magnetic field,
a coherent scattering line (such as a resonance line) produces linear
polarization following a dipole emission pattern (like that of a free
electron), but with an amplitude that depends on the details of the
particular transition.  In the presence of a relatively weak magnetic
field, the magnetic sublevels start to become non-degenerate in energy,
leading to an adjustment of the polarization
amplitude, which becomes a function of scattering direction with respect
to the local magnetic field vector (Stenflo 1994).

A description of this effect in terms of classical damped
harmonic oscillators is quite helpful because of its visual nature.
For simplicity, consider a level transition that has a polarization
amplitude of 100\% when scattering through a right angle, just like
Thomson scattering.  The scattering of unpolarized incident light,
typical of the case for illumination by starlight, is pictured
as the excitation of two orthogonal dipole oscillators.
Forward and backward scattered radiation is unpolarized.
                                                                                
Now consider a magnetic field that is perpendicular to the direction
of incident radiation.  The magnetic field exerts a Lorentz force
on the oscillating bound electron such as to precess the oscillation
about the axis of the field direction.  The competition here is
between the Larmor frequency $\omega_L = g_L B / m_{\rm e} c$ that
sets the rate of precession and the Einstein $A$-value that sets
the rate at which radiation is scattered.  For a small ratio of
$\omega_L / A$, precession is minimal, and the scattering is
essentially non-magnetic.  But when $\omega_L/A$ is large, precession
leads to a full rotation of the oscillator before much damping of
the amplitude occurs.  Consequently,
the scattered light when viewed along the
magnetic field becomes completely depolarized. We
refer to this limit as ``saturated'', because information about the
field strength is lost -- one knows the field is relatively strong,
but the low polarization is a hindrance for determining exactly how
strong, yet there is still information about the magnetic field
direction.  In terms of synthetic polarization spectra from models,
the saturated limit is valuable for interpreting the results because
of its simplistic properties -- complete depolarization along the
field, but no precession of the dipole oscillator that is parallel
to the field.  At its heart the Hanle effect is about redistributing
scattered light relative to the zero field case.

There have been a series of papers highlighting applications of the
Hanle effect to scattering lines from winds (Ignace et~al.\ 1997;
Ignace et~al.\ 1999; Ignace 2001a; Ignace 2001b; Ignace et~al.\ 2004).
These have dealt exclusively with line polarizations from optically thin
scattering.  Ignace et~al.\ (2004) consider the impact of line optical
depth on the polarization through a single-scattering approximation,
whereby optical depths below unity adopt the single-scattering results,
but zero polarization contributions are assumed from regions where the
optical depth exceeds unity.  The series has dealt with spherical wind
flows, expanding disks, and simplified considerations of oblique magnetic
rotators.  

\begin{figure}[H]
\begin{center}
\includegraphics[width=\columnwidth]{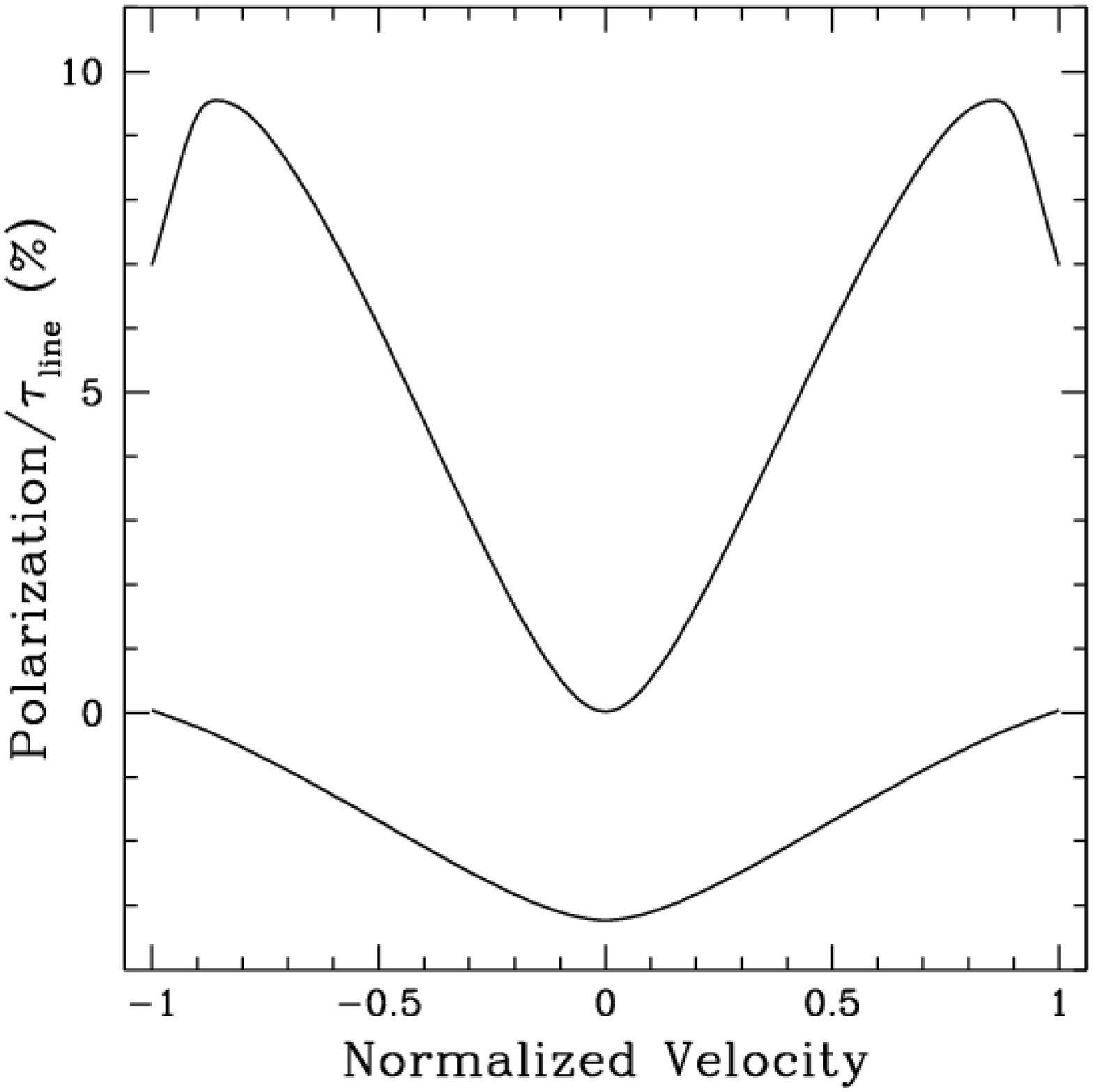}
\caption{The Hanle effect for a thin scattering line from a Keplerian disk
seen edge-on.  The polarization is normalized to the line optical
depth $\tau_{\rm line} < 1$.
The upper curve shows the polarization without a magnetic field.  Lower
curve is for a toroidal field in the saturated limit.  The sign change
signifies a rotation of the polarization position angle by $90^\circ$.
\label{ignace:fig1}}
\end{center}
\end{figure}

A new model presented at this meeting was a calculation for an optically
thin line from an axisymmetric Keplerian disk,
with results shown in Figure~\ref{ignace:fig1}.  The upper curve shows
the total flux emission profile, and the 
lower curve shows two curves for
the polarized line profile.  These are plotted against velocity shift
normalized to the Keplerian speed at the stellar radius.  In the lower
panel, the upper curve is the polarization without a magnetic field,
and the lower one is with a toroidal field, $B_\varphi \propto r^{-1}$,
in the saturated limit.  
Note that these profiles
assume an edge-on viewing perspective and normalized to
line optical depth, $\tau$.  Going to higher inclinations
can affect the profile shape, but the dominant effect is to lower the
amplitude of the polarization.  Also, this calculation does not take
account of absorption of the photospheric continuum, nor contamination by
photospheric lines, nor stellar occultation of the rearward disk (e.g.,
the approach of Ignace 2000).  However, with the disk velocity field being
right-left anti-symmetric, the effects will 
be symmetric about line-center.

Of particular interest is that the line-integrated polarization is
non-zero, and so even narrow-band polarimetry could be used in order to
increase signal-to-noise to detect the influence of the Hanle effect.
Different lines that are sensitive to different field
strengths would yield not only different levels of polarization, but
even net position angle rotations.  

There are plans to measure the Hanle effect for the first time in
stars other than the Sun.  The Far Ultraviolet Spectro-Polarimeter
(FUSP, see www.sal.wisc\\ .edu/FUSP; Nordsieck et~al.\ 2003; Nordsieck
\& Ignace 2005) will have the capability at a resolving power of
$R\approx 1800$ of measuring the linear polarization across wind-broadened
P~Cygni lines of bright stars.  This is a rocket payload mission expected
to have multiple flights.  The stars targeted for detecting the Hanle
effect are $\zeta$~Ori and $\xi$~Per in the missions second
launch, currently scheduled for late 2009.

\section{Zeeman Effect for Winds}

As is well known, the Zeeman effect describes how a magnetic field
leads to splitting of atomic sublevels.  In a standard Zeeman triplet,
one generally has an unshifted line component that can be linearly
polarized (referred to as a ``$\pi$'' component) and a pair of equally
shifted components left and right of line center range (referred to as
``$\sigma$'' components).  The $\sigma$ components are circularly
polarized when
viewed along the magnetic field, in which case the unshifted $\pi$
component will not be seen, but are linearly polarized when viewed
orthogonal to the magnetic field.

In the 
weak-field
limit -- not so weak as to be in the Hanle regime,
but sufficiently weak that the Zeeman splitting is small compared to
other broadening processes, the Zeeman components will be strongly
blended.  In the Hanle regime, the $\sigma$ components maintain a phase
relation, leading to linear polarization effects owing to the coherent
superposition of left and right circular polarizations.  In the weak
Zeeman regime, the $\sigma$ components are distinctly split relative to
their respective natural broadening, and the circular polarizations of
the two components add incoherently.  Consequently, blending from thermal,
turbulent, rotational, or wind broadening strongly diminishes the net
circular polarization of the line.

Ignace \& Gayley (2003) explored the Zeeman effect in the Sobolev
approximation in order to determine the scaling of the circular
polarization on magnetic field and wind properties.  In the context
of the longitudinal Zeeman effect, that relates to the net circular
polarization of a line, and scales with the net projected magnetic flux
along the line-of-sight, the circular polarization is derived from a
Taylor expansion of the difference in intensity between the two $\sigma$
components.  Following that paper, we define $I_\pm$ as left (blueshifted)
and right (redshifted) circularly polarized intensities as given by

\begin{equation}
I_\pm \approx \frac{1}{2}\,I_0 (\Delta \lambda\mp \Delta \lambda_B\,\cos \gamma),
\end{equation}
                                                                                
\noindent where $I_0$ is the intensity profile shape in the absence
of a magnetic field, $\Delta \lambda$ is the wavelength shift from
line center, and
                                                                                
\begin{equation}
\cos \gamma = \hat{B} \cdot \hat{z},
\end{equation}
\noindent for $\hat{B}$ the magnetic field unit vector and $\hat{z}$ a
unit vector directed toward the observer.  For the intensity of
circularly polarized light, we have Stokes $V = I_+ - I_-$, yielding

\begin{equation}
V = -  \Delta \lambda_B\,\cos \gamma\,\left(\frac{dI_0}{d\lambda}
        \right)_{\Delta \lambda}
\end{equation}

In the Sobolev approximation for spherical winds, one builds up a line
profile by considering isovelocity surface ``cuts'' through the wind
flow, and integrating the intensities across these surfaces, accounting
for stellar occultation and absorption of the photospheric continuum.
In the 
weak-field
regime of interest, the fluxes $F_\pm$ are identical
in shape but slightly shifted from one another.  The difference of these
profiles gives the flux of circular polarization $F_V$.
Focusing on only the emission for illustration, 
Ignace \& Gayley derive the formula:

\begin{eqnarray}
F_V^{\rm emis} (\Delta \lambda_{\rm z}) & = & -\frac{2\pi}{D^2}\,
        \int_{\Delta \lambda_{\rm z}}\, \Delta \lambda_B \, \cos\gamma\, \nonumber \\
 & &     \frac{d}{d\Delta \lambda_{\rm z}}\,\left[ S_\lambda \,
        \left ( 1-e^{-\tau_S} \right )\right] \, p \, dp,
        \label{ignace:Vemis}
\end{eqnarray}

\noindent where $\tau_S$ is the Sobolev optical depth,
$\Delta \lambda_{\rm z}$ identifies the wavelength
shift in the profile from line center that spatially corresponds to
an isovelocity zone, $S_\lambda$ is the position-dependent source
function, $D$ is the source distance, and $p$ the polar radius in
observer coordinates.  Implicit is that the wind is spherical, that
the field is axisymmetric, and that the viewer perspective is along the
field symmetry axis (otherwise there would be an integration in observer
azimuth $\alpha$ since the intersection of the field topology with the
isovelocity zones would not generally be azimuthally symmetric).  Consequently,
equation~(\ref{ignace:Vemis}) is maximized for the net magnetic flux
through isovelocity zones, and the resultant circular polarizations
represent best-case scenarios.

\begin{figure}[H]
\begin{center}
\includegraphics[width=\columnwidth]{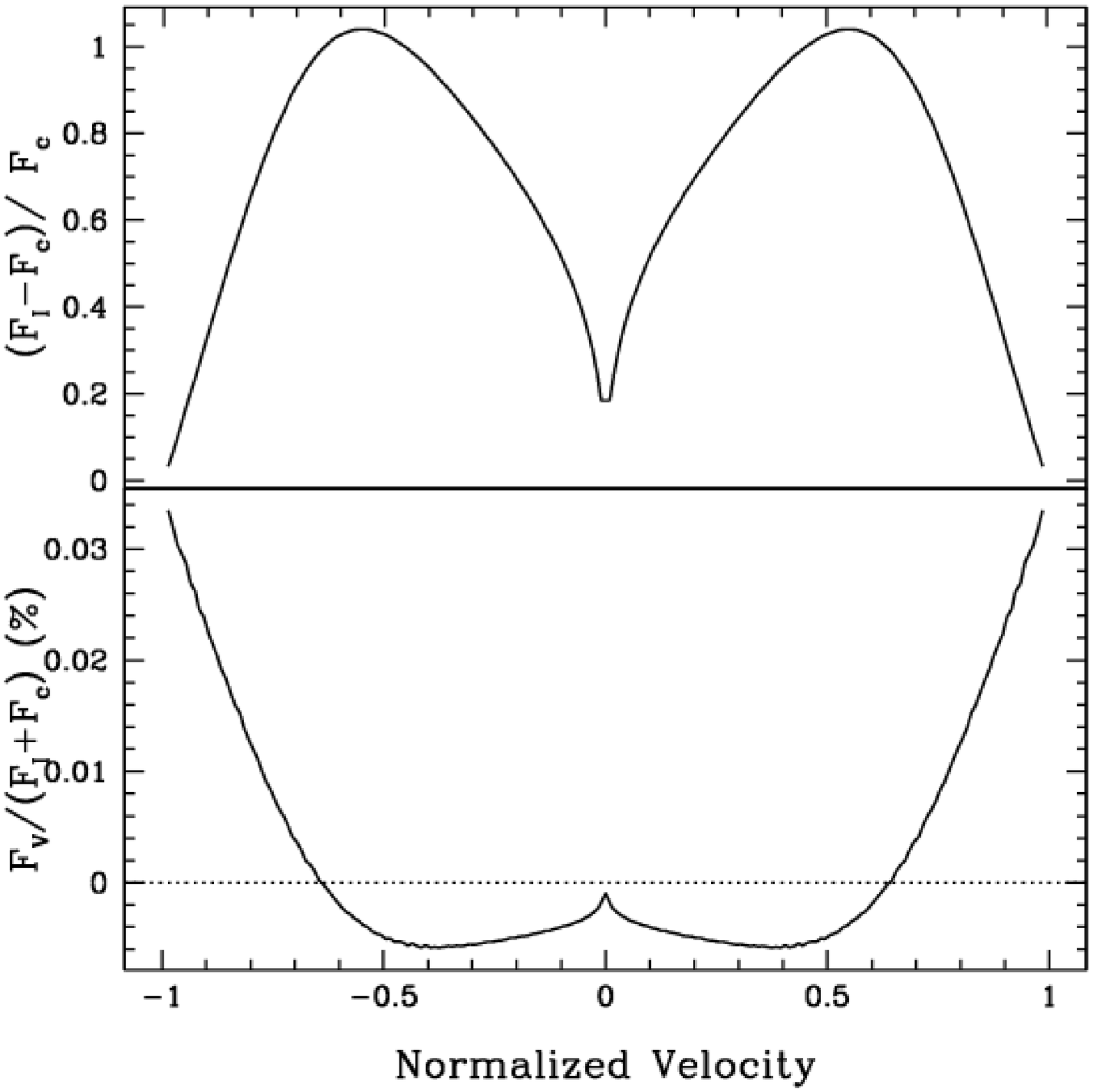}
\caption{Upper panel shows the emission line profile from a Keplerian disk
(relative to and normalized by the continuum level).  Lower panel displays
the percent circularly polarized profile assuming a toroidal magnetic field
with $B_\ast = 100$~G for an optical line.
\label{ignace:fig2}}
\end{center}
\end{figure}

Ignace \& Gayley derived polarized line profiles for simplified models
of resonance and recombination lines, assuming a velocity law
that was linear with radius and using simple field distributions such as
a split monopole.  As expected, the overall peak amplitude of the
polarization scales with $v_Z/v_\infty$, for $v_Z$ the velocity splitting
of the Zeeman components.  Peak polarizations of about 0.05\% were
found, assuming a surface field strength of 100~G and a modest terminal
wind speed of 1500 km/s.  Such values are challenging, but not
beyond the capability of existing and upcoming telescopes, such
as the Potsdam Echelle Polarimetric and Spectroscopic Instrument
(``PEPSI'', see www.aip.de/pepsi; Hofmann et~al.\ 2002; Andersen et~al.\
2005).

We are taking steps in developing approaches for computing line
polarization for more realistic stellar winds.  The two main practical
considerations are:  (1) what geometries are of most interest and (2)
what geometries are most observationally feasible?  The answers to
both questions would seem to be the same, namely circumstellar disks.
Disks in Keplerian (or near Keplerian) rotation are relatively common:
for example protostellar disks, interacting binaries, and Be~disks.
The scaling of polarization amplitude with $v_Z/v_{\rm max}$, where
$v_{\rm max}$ is the maximum flow speed in a system, is robust, and
Keplerian disks are limited by the speed of critical rotation of their
central star, which is typically a factor of 3 smaller than wind speeds.
Consequently, line polarizations
will be larger by a similar factor for a given surface field strength.

As an example, we consider a Keplerian disk with a purely toroidal
magnetic field.  As noted previously, the isovelocity zones
of an axisymmetric rotating disk are left-right symmetric, in contrast
to the back-front symmetry for spherically expanding winds.  For disks
the isovelocity zones are loops.  As for the calculation with the
Hanle effect, we focus on just the emission contribution and for now
ignore absorption and stellar occultation.  For modeling the disk 
emission, we follow the escape probability approach of Rybicki
\& Hummer (1983).

A useful notational device is to scale the Zeeman splitting to a
Doppler shift in the units of an equivalent line-of-sight speed $v_Z$,
and then the simplest preliminary result can be obtained assuming a
toroidal disk field obeying $B_\varphi \propto r^{-1/2}$.
Then the Zeeman shifts of the $\sigma_\pm$ components, characterized by $v_Z$, 
scale in direct proportion to the actual Keplerian speed $v_\phi$.
As a result, the isovelocity zones for
the respective $\sigma_\pm$ components are identical 
to Keplerian but with $v_\phi$ scaled {\it uniformly} larger 
or smaller, by $(v_\phi \pm v_Z)/v_\phi$, 
for each circular polarization.

In this special case, the resultant
Stoke-V flux for the line profile is given by the expression

\begin{eqnarray}
F_V^{\rm emis} & = & F_+^{\rm emis} - F_-^{\rm emis} \\
    & = & -2\,\left(\frac{v_B}{v_{\rm rot}}\right)\,
	\left[F_I + \Delta \lambda_{\rm z} \,\left(\frac{dF_I}
	{d\lambda}\right)\right],
\end{eqnarray}

\noindent where $v_B$ is $v_Z$ evaluated for a surface field strength at
the equator of the star with $B_\ast = 100$~G,
and $v_{\rm rot}=500$~km/s is the Keplerian rotation speed
at the radius of the star.  The resultant profile shape is shown in the
lower panel of Figure~\ref{ignace:fig2}, with the upper panel displaying
the Stokes $F_I$ profile for the line emission.  The profiles are plotted
against observed velocity shift normalized to $v_{\rm rot}$.  Note that
the $F_I$ profile shows the characteristic double-peak morphology,
whereas the polarized line shows its strongest values at the extreme
line wings.  The circularly polarized emission from a disk is seen to
be left-right {\it symmetric}, in contrast to a spherical wind that is
{\it antisymmetric} about line center.


\section{Observing Strategies}

We have emphasized what lies ahead for the future opportunities in direct
detection of circumstellar magnetic fields, in order to test models
of magnetized plasma flows.  It is significant that efforts in this
regard are already underway.  As mentioned, Donati et~al.\
(2005) have claimed a detection of magnetic fields in the circumstellar
disk of FU~Ori.  Hubrig et~al.\ (2007) have also claimed a detection
of circular polarization in the circumstellar line of a couple of Herbig
Ae stars; however, this has been contested by Wade et~al.\ (2006).
Eversberg et~al.\ (1999) and St-Louis et~al.\ (2007) have searched for the
Zeeman effect in lines of Wolf-Rayet stars, although they have no confirmed
detections as yet.  The key point is that observers are undertaking these
searches, albeit with difficulty.  More detections are to be expected,
so diagnostic procedures are needed to connect the data with models of
magnetized winds and disks.

This raises the obvious question, how are the Zeeman and Hanle effects to
be most effectively employed?  Bear in mind that the Hanle effect only
works for scattering lines, but it is sensitive to quite weak fields,
in the range 1--100~G.  For hot stars this generally relegates its
usefulness to UV spectropolarimetry, which of course requires space-borne
instrumentation.  Fortunately, FUSP should give us our first opportunity
of sampling the Hanle effect in the UV lines of hot stars.

On the other hand, there are limited classes of objects where even
H$\alpha$ can act as a scattering line, significant for the fact that it
can be observed from the ground, and sensitive to fields of around 1~G.
Such sources include yellow hypergiants and some blue supergiants (e.g.,
Verdugo et~al.\ 2005).  Another important class are supernovae, as for
example the polarization from H$\alpha$ seen in SN1987A (Jeffery 1987,
1991).  Studies of polarization in SNe suggests that observed variations
can arise in part from line scattering effects e.g., Hoffman 2006).
In those cases where the polarization arises from line scattering,
comparisons of the polarizations between different lines and in relation
to the continuum polarization from Thomson scattering could reveal the
presence of the Hanle effect and thereby constrain magnetic fields in
the ejecta of SNe.


The magneto-rotational instability (MRI -- Balbus \& Hawley 1991) has been
found to be a robust mechanism for producing turbulent magnetic fields.
In particular in a Keplerian disk, simulations indicate that for an
initially vertical field threading the disk, the MRI leads to two primary
field components:  one that is predominantly toroidal (like that of our
model profiles) and one that is turbulent or ``randomized''.  Moreover,
the toroidal field likely switches direction between the upper half disk
and the lower half.  So for the Zeeman effect, the oppositely directed
toroidal field essentially leads to net zero magnetic flux around
the disk 
for optically thin emission, 
and so would not produce observable circular polarization.
This would not be the case for the Hanle effect,
as the result
shown in Figure~\ref{ignace:fig1} does not depend on the handedness
(or reversals) of the toroidal field in the disk.

A distinct advantage of the Hanle effect in {\it turbulent} magnetic 
regions is that it is not canceled by line-of-sight magnetic
field reversals, the way the longitudinal Zeeman effect is.
Indeed, the Hanle effect has been employed as a
diagnostic of turbulent solar magnetic fields (Stenflo 1982;
Stenflo et~al.\ 1998).  (Note however that for an unresolved source, a
field that is tangled on a spatial scale that is small compared to the
Sobolev length will likely lead to complete depolarization from that
region.)  

Perhaps the best strategy is to employ the Zeeman and Hanle effects in a
complementary fashion.  The Hanle effect will likely be best sensitive to
weak fields from scattering lines in regions where the line is optically
thin (Ignace et~al.\ 2004), even if the surface field is quite strong,
because the circumstellar field will typically drop rather rapidly
with radius (as for multipole fields).  The Zeeman effect will be
sensitive to strong photospheric fields, and possibly circumstellar
fields in the inner wind or disk.  Both of these should be used along
with additional sources of information about the source, such as the
continuum polarization that may arise from electron scattering, and line
profile shapes in Stokes-$F_I$.

We suggest that one promising target for honing these diagnostics is
$\sigma$~Ori~E.  This Bp~star has a Zeeman detection (Landstreet \&
Borra 1978), has anomalous X-ray behavior (Groote \& Schmitt 2004),
and cyclic variations in its H$\alpha$ emission (Townsend et~al.\
2005), all that been successfully interpreted in terms of a strongly
magnetized circumstellar envelope (Townsend \& Owocki 2005).  This and
similar sources where the magnetic field properties are already highly
constrained would be good targets for detecting the Zeeman and Hanle
effects in circumstellar lines. \\

The authors would like to thank Ken Nordsieck for discussions of the
Hanle effect in SNe, and Jennifer Hoffman for a preview of recent
line polarization data in SNe.

\end{document}